\documentclass[APS,twocolumn,floatfix]{revtex4}
\usepackage{graphicx}
\usepackage{calc}
\usepackage{pifont}
\usepackage{amssymb}
\usepackage{epsfig}
\usepackage{amssymb}
\usepackage{psfig}
\usepackage{amsmath}
\begin{document}

\title{Fabrication and Characterization of Electrostatic Quantum Dots in
a Si/SiGe 2D Electron Gas, Including an Integrated Read-out Channel}

\author
{M. R. Sakr $^{1}$, E. Yablonovitch$^{2}$, E. T. Croke$^{3}$, and H. W. Jiang$^{1} $}

\affiliation{$ ^{1}$Department of Physics, University of California, Los Angeles, Los Angles, California 90024\\
             $ ^{2}$Department of Electrical Engineering, University of California, Los Angeles, Los Angles, California 90024\\
             $ ^{3}$HRL Laboratories, LLC, 3011 Malibu Canyon Rd RL63, Malibu, CA 90265}
\begin{abstract}
A new fabrication technique is used to produce quantum dots with read-out channels in silicon/silicon-germanium two-dimensional electron gases. The technique utilizes Schottky gates, placed on the sides of a shallow etched quantum dot, to control the electronic transport process. An adjacent quantum point contact gate is integrated to the side gates to define a read-out channel and thus allow for noninvasive detection of the electronic occupation of the quantum dot.  Reproducible and stable Coulomb
oscillations and the corresponding jumps in the read-out channel resistance are observed at low temperatures. The fabricated dot combined with the read-out channel represent a step towards the spin-based quantum bit in Si/SiGe heterostructures.
\end{abstract}

\maketitle

Individual spins in silicon/silicon-germanium (Si/SiGe) heterostructures have many desirable properties \cite{Vrijen2000} for solid-state implementations of spin-based quantum information processing.   In silicon, electrons can have very long coherence times since they experience a very weak spin-orbit coupling and  zero hyperfine interaction to nuclear spins in isotopically-purified structures \cite{Tyryshkin2003}.   The tunable spin-orbital coupling and the ability to control the electron
wave-functions allow the execution of electrically controlled g-factor tuning for logic operations \cite{Vrijen2000}.

Despite its potentials, the technologies for fabricating submicron devices in Si/SiGe two-dimensional electron gases (2DEGs) are not as established as those for GaAs/AlGaAs materials.  One main technical difficulty is that the surface Schottky gates cannot be reliably made due to problems of Fermi level pinning and leakage paths to the 2DEG created by dislocations in strained materials.  To overcome this difficulty, X. Z. Bo \textit{et al.} \cite{Bo2002}  and L. J. Klein \textit{et al.}
\cite{Klein2004}  have recently fabricated quantum dots (QDs) by using atomic force microscope lithography on Si/SiGe heterostructures.  In their devices, trenches are created by the lithography and the isolated 2D electron regions are used as gates to control a QD surrounded by the trenches.

Here we demonstrate a different fabrication technique to produce QDs in a modulation-doped Si/SiGe heterostructure.  Our method has a better gate-QD coupling allowing for efficient electrostatic control of the QD.  The QD can be potentially squeezed to hold only one electron or, in other words, a single spin, which is of particular interest for the purposes of quantum computation.  Furthermore, a read-out channel can be fabricated adjacent to the dot for a noninvasive read-out of the spin state of the
QD, similar to other GaAs/AlGaAs devices \cite{Elzerman2004}.

Molecular beam epitaxy (MBE) was used to grow the Si/SiGe heterostructure. The 2DEG resides in a 10 nm silicon channel 21.5 nm below the surface.  Antimony (Sb) is incorporated into the donor layer via a delta-doping approach using an Sb$_4$ source. The donor layer is separated from the silicon channel by a 5 nm Si$_{0.7}$Ge$_{0.3}$ spacer layer. A 10 nm Si$_{0.7}$Ge$_{0.3}$ layer and 6.5 nm Si cap were grown on the spacer layer. From the substrate to the Si channel, a step-graded SiGe/SiGeC buffer sequence and a 400 nm Si$_{0.7}$Ge$_{0.3}$ layer were grown.
At room temperature, the mobility and concentration of the 2DEG are 1700 cm$^2$/Vs and $1.1 \times 10^{12}$ cm$^{-2}$ respectively. At 4 K, these two quantities measure $1.2 \times 10^{4}$ cm$^{2}$/Vs and $1.2 \times 10^{12}$ cm$^{-2}$  respectively. Aluminum ohmic contacts to the 2DEG are established using rapid thermal annealing at $420 ^{\circ}C$ for 2 min.

\begin{figure}[htb]
\scalebox{1}{\includegraphics[width=8.cm,height=9 cm,angle=0]{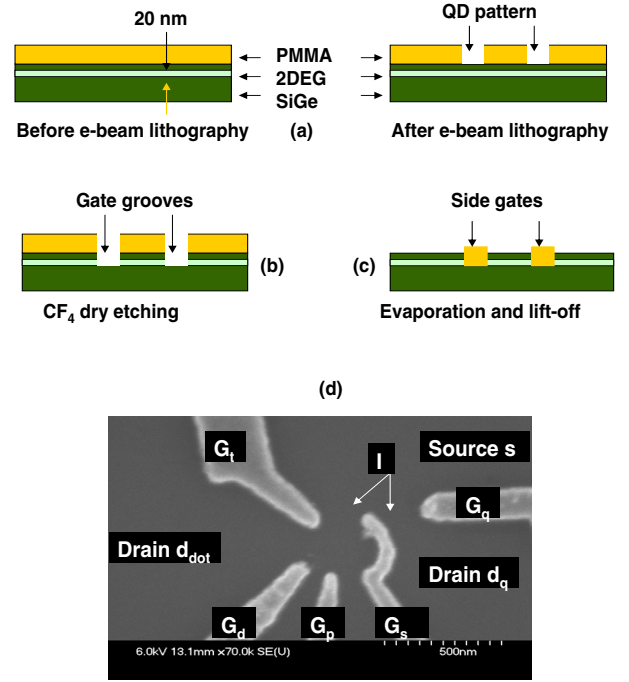}}
\caption{(a)-(c)  Steps for fabricating the side-gated structure (described in detail in the text). (d) Scanning electron micrograph of the structure.  Gates $G_s$, $G_d$ and $G_t$ define the quantum dot, while $G_p$ is a plunger gate used to vary the number of electrons inside the dot. $G_q$ forms a QPC that senses the electronic occupation of the dot. }
\end{figure}

In our device, metallic side-gates are used similarly to that of the vertical QD in GaAs/AlGaAs heterostructures \cite{Austing1995}.   The steps used to produce the Schottky gates are shown in Fig. 1.  Electron beam lithography is used to pattern the QD structures on PMMA, Fig. 1(a). After developing, CF4 reactive ion etching is carried out to remove the 2D electron part of the device structure and to establish grooves for the side gates as shown in Fig. 1(b). The etching was very shallow reaching about 30 nm underneath the surface of the sample. This process is followed by evaporating a 10 nm/70 nm titanium/gold layer as indicated in Fig. 1(c), then the QDs (Fig. 1(d)) are obtained by metal lift-off in acetone. The etching process results in side-wall depletion, which insulates the metallic gates from the electron gas. Since titanium forms a Schottky barrier with the side SiGe wall, we expect relatively fewer dangling-bond-induced  traps compared to that in unpassivated trenches \cite{Bo2002,Klein2004}.
We have tested the leakage through the Schottky side gates to the 2DEG and found that most gates leak at positive gate voltages around 0.4 V or higher and at negative gate voltages around -3.0 V or lower.

\begin{figure}[htb]
\scalebox{1}{\includegraphics[width=8.cm,height=10 cm,angle=0]{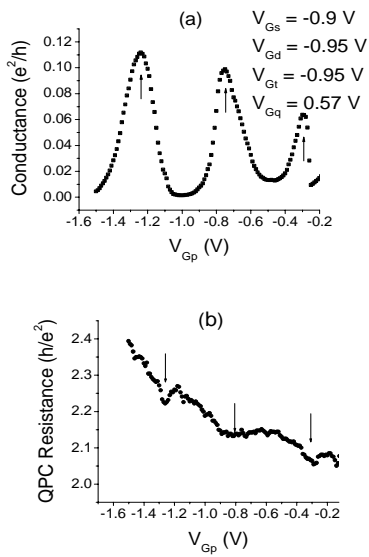}}
\caption{Coulomb oscillations and the corresponding dips in the channel resistance.
(a) The peaks are periodic in $V_{Gp}$ with a period of $\Delta V_{Gp} \approx$  0.45 V.  (b) Each dip in the QPC channel resistance corresponds to a change in the number of electrons on the dot by one
 }
\end{figure}

Now we shall demonstrate some of the characteristic features of the fabricated structures.  Figure 2(a) shows Coulomb oscillations in the conductance of a quantum dot as a function of the plunge gate voltage $V_{Gp}$ at temperature $T = 4.2$ K keeping other gate voltages at constant values. The conductance through the QD is measured using a standard lock-in technique with an ac voltage of a few hundred microvolts between the source and drain to avoid self-heating of the QD.  In the figure, the
number of electrons in the dot is decreased one-by-one when the magnitude of the voltage on the gate is increased.  Devices with different sizes show a reproducible single electron charging effect and are stable over a few hours.

For quantum information processing, it is undesirable to measure the spin state by passing current through the dot.  We have used a side quantum point contact (QPC) as an electrometer for noninvasive sensing of the number of electrons in QD, as this discrete electronic occupation can be used for projective spin-state read-out \cite{Elzerman2004}.  In the tunneling regime, the resistance of the QPC channel is very sensitive to the potential variations caused by adding or expelling electrons from the QD
\cite{Field1993}.   Figure 2(b) shows that there are clear futures in the QPC channel resistance corresponding to the peaks in the Coulomb oscillations in Fig. 2(a).  The spacing between two successive peaks is periodic and equals $\Delta V_{Gp} \approx  0.45$ V for that particular quantum dot.

\begin{figure}[htb]
\scalebox{1}{\includegraphics[width=8.cm,height=10 cm,angle=0]{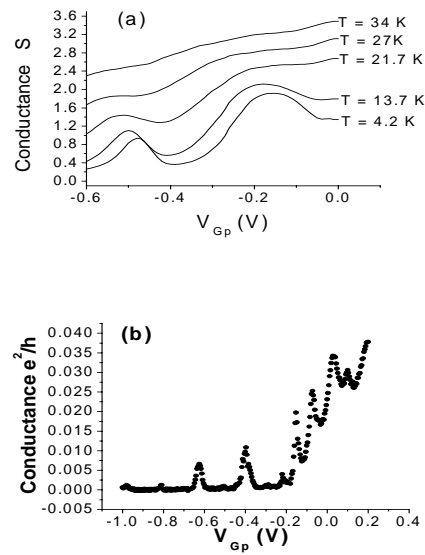}}
\caption{(a) Coulomb oscillations at different temperatures for a small dot.  The peaks survive up to 34 K.  (b) Coulomb oscillations at 0.4 K for a larger dot.}
\end{figure}

The evolution of the Coulomb oscillations for a small dot (geometrical diameter about 200 nm) at different temperatures is shown in Fig. 3(a). As the temperature increases, the oscillations broaden and they finally disappear at $T = 34$ K.  This rather large Coulomb charging energy is a result of the small QD size.  Figure 3(b) shows the Coulomb oscillations for a larger dot (geometrical diameter about 350 nm) at 0.4 K.  For the larger dot, the oscillations become visible at about $T < 4$ K.  Larger
number of oscillations, and smaller $\Delta V_{Gp}$ are also seen, as expected.

The single electron transport is also manifested in a so-called stability diagram that represents the differential conductance $dI_{dot}/dV_{sd}$ of the QD measured at different values of the source-drain voltages and the gate voltages on the plunge gate. A stability plot, obtained for the larger dot at 0.4 K with $125$ $\mu$V ac excitation voltage applied between source and drain at 15 Hz, is shown in Fig. 4.  The charging energy $E_C$ equals to $e\Delta V_{Gp}(dV_{sd}/dV_{Gp})/2$ and can be inferred
from the figure by calculating the slope of the ``diamonds" \cite{Klein2004}. The charging energy of 2 meV corresponds to a total capacitance $C$ of 80 aF.  Representing the dot as a metallic disc with radius R, the total capacitance of the dot reads $C = 8\epsilon_0\epsilon_rR$ giving an effective diameter of about 190 nm for the QD.  Here, $\epsilon_0$   is the permittivity of vacuum and $\epsilon_r$ is the dielectric constant of silicon.

\begin{figure}[htb]
\scalebox{1.}{\includegraphics[width=8.cm,height=6 cm,angle=0]{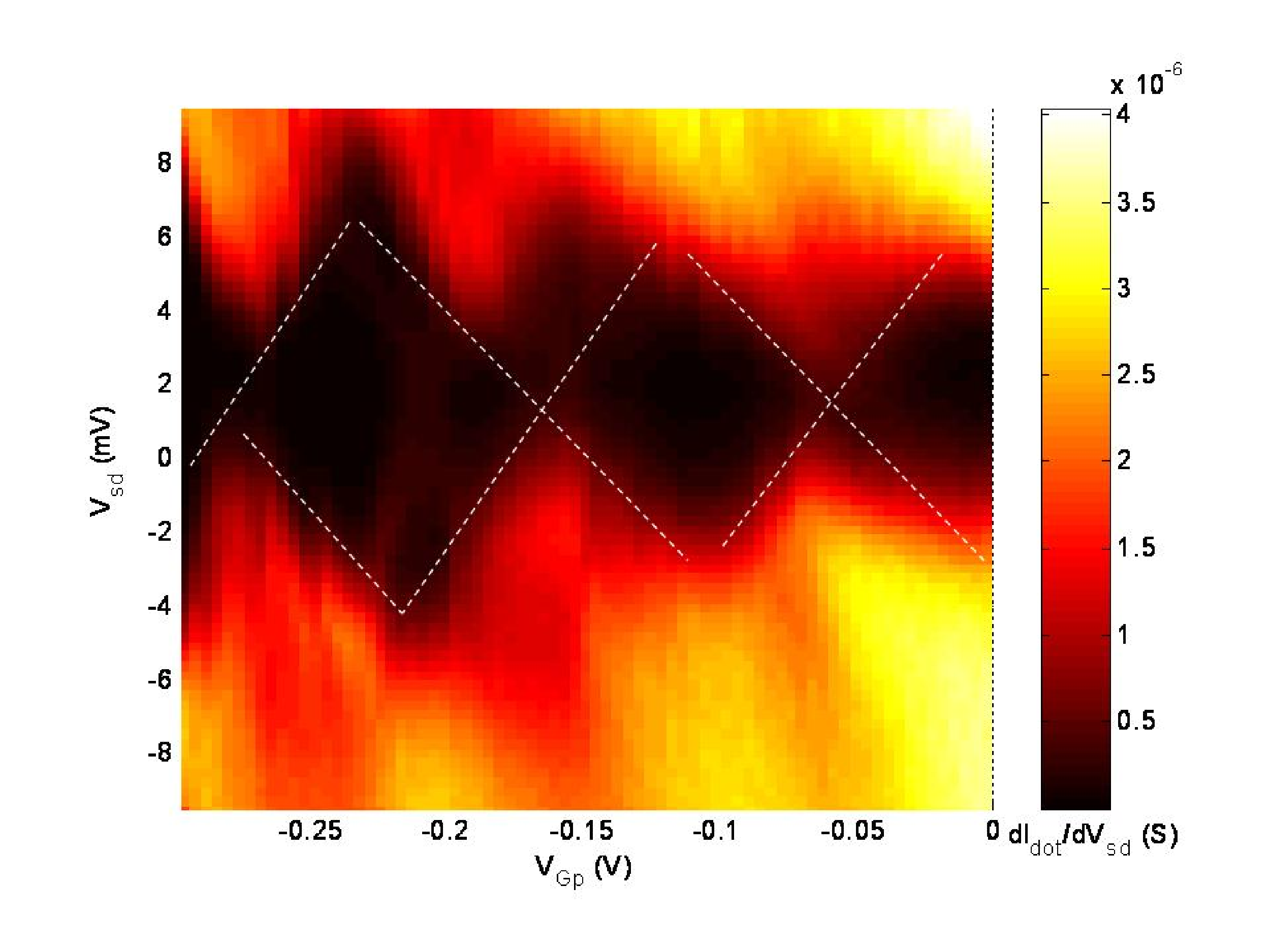}}
\caption{Stability plot of the differential conductance in a gray scale as a function of the source-drain voltage $V_{sd}$ and the plunge gate voltage $V_{Gp}$ at 0.4 K.}
\end{figure}

In general, we are able to vary the number of electrons in the QDs one by one from 30-150 electrons.  The number of electrons in the smallest dot with an effective diameter of about 75 nm still has a minimum of 30 electrons.  This minimum number is limited by the maximum gate voltage which can be used without leakage.  Currently, we are fabricating devices  using another Si/SiGe heterostructure that has a much lower 2D electron concentration (about $3 \times 10^{11}$
cm$^{-2}$, a factor of 6 smaller than that of the current heterostructure) and a higher mobility of $8 \times 10^{4}$ cm$^2$/Vs.  With these new devices, we  expect to get into the few-electron regime, which becomes useful for quantum information applications.

In summary, a quantum dot with an integrated charge read-out channel has been fabricated using a 2DEG in a silicon/silicon-germanium heterostructure.  The QD dot is laterally confined by metallic side-gates in etched grooves.  The devices show a reproducible single electron charging effect that is stable over an extended period of time. The discrete electronic occupation of the quantum dots is effectively detected using the adjacent QPC electrometer.  This
device structure with improvement can be potentially used as a spin-based qubit. This work is supported by the Defense MicroElectronics Activity under grant number DMEA 90-02-2-0217 and MARCO MSD Center.

\end{document}